\documentclass[aps,preprintnumbers,showpacs,superscriptaddress]{revtex4}
\usepackage{amssymb}
\usepackage{amsmath}
\usepackage{amsopn}
\usepackage{graphics}
\usepackage{graphicx}
\usepackage{epsfig}

\begin{document}

\title{Equilibrium statistical mechanics for single waves and wave spectra in Langmuir wave-particle interaction}

\author{M.-C. Firpo}
\affiliation{Laboratoire de Physique et Technologie des Plasmas (CNRS UMR 7648), Ecole
Polytechnique, 91128 Palaiseau cedex, France}
\author{F. Leyvraz}
\affiliation{Centro de Ciencias Fisicas, UNAM, Av. Universidad s/n,
Col. Chamilpa, C. P. 62210, Cuernavaca, Morelos, Mexico}
\author{G. Attuel}
\affiliation{Laboratoire de Physique et Technologie des Plasmas
(CNRS UMR 7648), Ecole Polytechnique, 91128 Palaiseau cedex, France}
\date{\today }
\preprint{}

\begin{abstract}
Under the conditions of weak Langmuir turbulence, a self-consistent
wave-particle Hamiltonian models the effective nonlinear interaction
of a spectrum of $M$ waves with $N$ resonant out-of-equilibrium tail
electrons. In order to address its intrinsically nonlinear
time-asymptotic behavior, a Monte Carlo code was built to estimate
its equilibrium statistical mechanics in both the canonical and
microcanonical ensembles. First the single wave model is considered
in the cold beam/plasma instability and in the O'Neil setting for
nonlinear Landau damping. O'Neil's threshold, that separates nonzero
time-asymptotic wave amplitude states from zero ones, is associated
to a second order phase transition. These two studies provide both a
testbed for the Monte Carlo canonical and microcanonical codes, with
the comparison with exact canonical results, and an opportunity to
propose quantitative results to longstanding issues in basic
nonlinear plasma physics. Then the properly speaking weak turbulence
framework is considered through the case of a large spectrum of
waves. Focusing on the small coupling limit, as a benchmark for the
statistical mechanics of weak Langmuir turbulence, it is shown that
Monte Carlo microcanonical results fully agree with an exact
microcanonical derivation. The wave spectrum is predicted to
collapse towards small wavelengths together with the escape of
initially resonant particles towards low bulk plasma thermal speeds.
This study reveals the fundamental discrepancy between the long-time
dynamics of single waves, that can support finite amplitude steady
states, and of wave spectra, that cannot.
\end{abstract}

\maketitle








\section{Introduction}

Wave-particle interaction is a universal phenomenon in space and
laboratory plasmas. Its importance ranges from magnetic confinement
fusion devices, in particular with its relation to anomalous
transport or supplementary heating, to laser plasma interaction,
astrophysics and charged particle beam physics. It is responsible
for high-frequency plasma turbulence which differs substantially
from low-frequency fluid or magnetohydrodynamic turbulence. In this
article, we shall consider one of the simplest wave-particle
interaction settings, namely electrostatic wave-particle interaction
taking place in bump-on-tail or weak beam-plasma instabilities.

Under the conditions of weak Langmuir turbulence \cite{Sagdeev},
electrostatic wave-particle interaction may be modeled \cite%
{Mynick78,CaryDoxas93,TMM,AEE98,book} by a reduced system coupling
self-consistently a set of $M$ longitudinal (Langmuir) waves to $N$
quasi-resonant tail particles. This approach is especially
appropriate to the description of basic plasma kinetic phenomena
such as the bump-on-tail instability \cite{discrete} or Landau
damping \cite{phasetransition}. In
this reduced framework, the effective dynamics is ruled by a $(M+N)$%
-dimensional Hamiltonian system. More specifically, particles in the
background plasma (whose velocities are roughly smaller than the thermal
velocity) only participate to the dynamics via a small subset of collective
modes, the Langmuir (long-wavelength) modes of longitudinal oscillation
around their guiding-center positions. These modes are described by
action-angle variables $\left( I_{j},\theta _{j}\right) $ with $1\leq j\leq
M $ for the $M$ waves. In the absence of resonant particles, they oscillate
with constant angular frequencies $d\theta _{j}/dt=\omega _{0j}$, that,
according to the Bohm-Gross dispersion relation, are approximately equal to
the plasma frequency $\omega _{p}$ for long-wavelength modes. These waves
strongly interact with those plasma particles in the tails of the
distribution function having velocities $v$ close to $\omega _{0j}/k_{j}$.
The coupling is controlled by the small parameter $\eta $ that is the ratio
of the tail density over the bulk plasma density. The dynamics of $N$
identical quasi-resonant particles moving on the interval of length $L$ with
periodic boundary conditions, with unit mass and charge, and respectively
position $x_{r}$ and momentum $p_{r}$, interacting with $M$ waves with
wavenumbers $k_{j}=j2\pi /L$, derives then from the Hamiltonian
\begin{equation}
H^{N,M}=\sum\limits_{l=1}^{N}\frac{p_{l}^{2}}{2}+\sum_{j=1}^{M}\omega
_{0j}I_{j}-\sqrt{\frac{2\eta }{N}}\sum\limits_{l=1}^{N}\sum_{j=1}^{M}\frac{%
\omega _{0j}^{3/2}\sqrt{I_{j}}}{k_{j}}\cos \left( k_{j}x_{l}-\theta
_{j}\right) .  \label{hamiltonian}
\end{equation}%
Equations of motion are then%
\begin{eqnarray}
\dot{x}_{l} &=&p_{l},  \label{eq_motion_1} \\
\dot{p}_{l} &=&-\sqrt{\frac{2\eta }{N}}\sum_{j=1}^{M}\omega _{0j}^{3/2}\sqrt{%
I_{j}}\sin \left( k_{j}x_{l}-\theta _{j}\right) ,  \label{eq_motion_2} \\
\dot{\theta}_{j} &=&\omega _{0j}+\frac{\omega _{0j}^{3/2}}{2}\sqrt{\frac{%
2\eta }{NI_{j}}}\sum\limits_{l=1}^{N}\cos \left( k_{j}x_{l}-\theta
_{j}\right) ,  \label{eq_motion_3} \\
\dot{I}_{j} &=&\sqrt{\frac{2\eta }{N}}\sum\limits_{l=1}^{N}\frac{\omega
_{0j}^{3/2}\sqrt{I_{j}}}{k_{j}}\sin \left( k_{j}x_{l}-\theta _{j}\right) .
\label{eq_motion_4}
\end{eqnarray}%
It is important here to insist on the principle of this derivation:
the background plasma is assumed to respond linearly to the waves
which is valid provided resonant particles, whose velocities are
much larger than plasma thermal velocity, only form a small
fraction, $\eta$, of the total plasma. Moreover it is interesting to
note that the same kind of wave-particle reduction has been
undertaken for other physical regimes of wave-particle interaction.
In particular, a Hamiltonian self-consistent wave-particle model
\cite{KRAFFT05} has recently been built in order to study the
nonlinear interaction of a packet of waves with a nonequilibrium
electron distribution in a magnetized background plasma.

Under the above hypotheses, the Hamiltonian model (\ref{hamiltonian})
contains the effective dynamics of weak Langmuir turbulence. Yet, if a
Hamiltonian system exhibits some ergodicity, then thermodynamics states that
time averages can be replaced with equivalent space averages over the
microcanonical ensemble. A Monte Carlo code was built to estimate the
equilibrium statistical mechanics of (\ref{hamiltonian}) in both the
canonical and microcanonical ensembles. This provides an efficient tool to
investigate the time-asymptotic fate of the wave-particle model and discuss
the evolution of weak Langmuir turbulence. This is of particular interest in
the broad spectrum case where direct thermodynamical derivations are
difficult.

In Section \ref{sec-mecastat-cano}, we present the canonical ensemble of the
$M$ waves/$N$ particles self-consistent Hamiltonian model and focus on the $%
M=1$ case. In Sec. \ref{sec-coldbeam}, we consider the cold beam/plasma
instability where the single wave approximation is particularly appropriate.
Then, in Sec. \ref{sec-ONeil-threshold}, we apply the single wave model to
the original O'Neil setting \cite{ONeil1965} for nonlinear Landau damping
and, in parallel with microcanonical Monte Carlo results, associate O'Neil's
threshold to a second order phase transition. These two studies provide both
a testbed for the Monte Carlo canonical and microcanonical codes, with the
comparison with exact canonical results, and an opportunity to propose
quantitative results to longstanding issues in basic nonlinear plasma
physics. It is shown that statistical mechanics is a useful quantitative
tool to address the nonlinear time-asymptotic saturation stage. Quantitative
comparisons with experiments are given. Then, in Sec. \ref%
{sec-small-coupling}, we turn to the properly speaking weak turbulence
framework by considering the case of a large spectrum of waves. We focus on
the small coupling limit, as a benchmark for the statistical mechanics of
weak Langmuir turbulence. We show that Monte Carlo microcanonical results
fully agree with an exact microcanonical derivation. We conclude by
discussing the predicted time-asymptotic spectrum, its relation to
quasilinear theory and to a pending debate in theoretical, experimental and
numerical plasma physics on the long time fate of the electric field for
Landau damping.

\section{Canonical ensemble}

\label{sec-mecastat-cano}

We shall start deriving statistical mechanics in the canonical ensemble and
then use the equivalence between canonical and microcanonical ensembles.
Apart from the energy, it can be easily checked that the average total
momentum per particle, $\sigma $, is also conserved with%
\begin{equation}
\sigma =p+\frac{1}{N}\sum_{j=1}^{M}k_{j}I_{j},  \label{totalmomentum}
\end{equation}%
and
\begin{equation}
p=\frac{1}{N}\sum_{l=1}^{N}p_{l}.  \label{sump}
\end{equation}%
We introduce the new particle momenta $\bar{p}_{l}=p_{l}-p$ and the reduced
wave intensities
\begin{equation}
\psi _{j}=\frac{I_{j}}{N},  \label{wave_intensities}
\end{equation}%
that are properly normalized with respect to the $N\rightarrow\infty$ limit
\cite{kinetic98}. Then, introducing (\ref{totalmomentum}), (\ref{hamiltonian}%
) reads
\begin{equation}
H=\sum\limits_{l=1}^{N}\frac{\bar{p}_{l}^{2}}{2}+\frac{N}{2}\left( \sigma
-\sum_{j=1}^{M}k_{j}\psi _{j}\right) ^{2}+N\sum_{j=1}^{M}\omega _{0j}\psi
_{j}-\sum\limits_{l=1}^{N}\sum_{j=1}^{M}\sqrt{2\eta \psi _{j}}\omega
_{0j}^{3/2}\cos \left( k_{j}x_{l}-\theta _{j}\right) .
\end{equation}%
The canonical partition function is $Z_{c}(N,\beta )=Z_{0}(N,\beta
)Z_{1}(N,\beta ,\sigma )$, with%
\begin{eqnarray}
Z_{0}(N,\beta ) &=&\int \prod\limits_{l=1}^{N}d\bar{p}_{l}\exp \left( -\beta
\sum\limits_{l=1}^{N}\frac{\bar{p}_{l}^{2}}{2}\right) \delta \left(
\sum_{l=1}^{N}\bar{p}_{l}\right) =\frac{1}{\sqrt{N}}\left( \frac{2\pi }{%
\beta }\right) ^{(N-1)/2},  \label{Z_zero} \\
Z_{1}(N,M,\beta ,\sigma ) &=&N^{M}L^{N}\int \prod\limits_{j=1}^{M}d\psi
_{j}d\theta _{j}\exp \left[ -N\beta g(\beta ,\sigma ,\mathbf{\psi },\mathbf{%
\theta )}\right] ,  \label{Z_1}
\end{eqnarray}%
where%
\begin{equation}
g(\beta ,\sigma ,\mathbf{\psi },\mathbf{\theta )=}\frac{1}{2}\left( \sigma
-\sum_{j=1}^{M}k_{j}\psi _{j}\right) ^{2}+\sum_{j=1}^{M}\omega _{0j}\psi
_{j}-\beta ^{-1}\ln \mathcal{I}(\beta ,\mathbf{\psi },\mathbf{\theta }),
\end{equation}%
and%
\begin{equation}
\mathcal{I}(\beta ,\mathbf{\psi },\mathbf{\theta })=\frac{1}{L}%
\int_{0}^{L}dx\exp \left( \beta \sqrt{2\eta }\sum_{j=1}^{M}\frac{\omega
_{0j}^{3/2}\sqrt{\psi _{j}}}{k_{j}}\cos \left( k_{j}x-\theta _{j}\right)
\right) .  \label{def_I}
\end{equation}%
All the difficulty lies in the estimation of (\ref{def_I}) for $M>1$. We
shall then first restrict our discussion to the case where only one mode is
selected before going to the large-$M$ case in the small-coupling limit in
Sec. \ref{sec-small-coupling}. The $M=1$ case occurs naturally in the cold
beam-plasma interaction \cite{ONeil1965,ONeil71,TMM,delCastillo2000}. We
get, for $k=1$ and $L=2\pi $,%
\begin{equation}
\mathcal{I}(\beta ,\psi ,\theta )=\frac{1}{2\pi }\int_{0}^{2\pi }dx\exp
\left( \beta \omega _{0}^{3/2}\sqrt{2\eta \psi }\cos \left( x-\theta \right)
\right) =I_{0}\left( \beta \omega _{0}^{3/2}\sqrt{2\eta \psi }\right) .
\end{equation}%
This gives%
\begin{equation}
g(\beta ,\sigma ,\psi ,\theta )=\frac{1}{2}\left( \sigma -\psi \right)
^{2}+\omega _{0}\psi -\beta ^{-1}\ln \left[ I_{0}\left( \beta \omega
_{0}^{3/2}\sqrt{2\eta \psi }\right) \right] .
\end{equation}%
Note here that in this particular case $g(\beta ,\sigma ,\psi ,\theta )$ is
actually independent of $\theta $. To perform the $\psi $ integral, we use
Laplace's method, for which we need to find the value $\psi ^{\ast }$ of $%
\psi $ for which $g(\beta ,\sigma ,\psi ,\theta )$ becomes a minimum. We get
\begin{equation}
\frac{\partial g}{\partial \psi }=\psi +\omega _{0}-\sigma -\frac{\eta \beta
\omega _{0}^{3}}{\beta \omega _{0}^{3/2}\sqrt{2\eta \psi }}\frac{I_{1}\left(
\beta \omega _{0}^{3/2}\sqrt{2\eta \psi }\right) }{I_{0}\left( \beta \omega
_{0}^{3/2}\sqrt{2\eta \psi }\right) }=0
\end{equation}%
for%
\begin{equation}
\psi ^{\ast }=\sigma -\omega _{0}+\eta \beta \omega _{0}^{3}\frac{%
I_{1}\left( \beta \omega _{0}^{3/2}\sqrt{2\eta \psi ^{\ast }}\right) }{\beta
\omega _{0}^{3/2}\sqrt{2\eta \psi ^{\ast }}I_{0}\left( \beta \omega
_{0}^{3/2}\sqrt{2\eta \psi ^{\ast }}\right) }.  \label{identity_phistar}
\end{equation}%
We have, $\forall \psi \geq 0$, $\partial g/\partial \psi \geq \left.
\partial g/\partial \psi \right\vert _{0}=\omega _{0}-\sigma -\eta \beta
\omega _{0}^{3}/2$. If $\omega _{0}-\sigma -\eta \beta \omega _{0}^{3}/2\geq
0$, $g$ attains its minimum at the boundary, namely in $\psi =0$. Otherwise,
there exists a unique minimum in $\psi ^{\ast }>0$ satisfying (\ref%
{identity_phistar}). Transition occurs for the critical
\begin{equation}
\beta _{c}=\frac{2\left( \omega _{0}-\sigma \right) }{\eta \omega _{0}^{3}}.
\label{beta_critical}
\end{equation}%
We have then%
\begin{eqnarray*}
Z_{1}(N,\beta ,\sigma ) &=&2\pi N\int_{0}^{\infty }d\psi \exp \left[ -N\beta
g(\beta ,\sigma ,\psi \mathbf{)}\right] \\
&\simeq &2\pi N\exp \left[ -N\beta g(\beta ,\sigma ,\psi ^{\ast }\mathbf{)}%
\right] \int_{0}^{\infty }d\psi \exp \left[ -\frac{N\beta }{2}\left. \frac{%
\partial ^{2}g}{\partial \psi ^{2}}\right\vert _{\psi ^{\ast }}(\psi -\psi
^{\ast })^{2}\right] \\
&=&\pi N\exp \left[ -N\beta g(\beta ,\sigma ,\psi ^{\ast }\mathbf{)}\right]
\sqrt{\frac{2\pi }{N\beta \left. \frac{\partial ^{2}g}{\partial \psi ^{2}}%
\right\vert _{\psi ^{\ast }}}}\left[ 1+\mathop{\mathrm{erf}}\left( \sqrt{%
\frac{N\beta }{2}\left. \frac{\partial ^{2}g}{\partial \psi ^{2}}\right\vert
_{\psi ^{\ast }}}\psi ^{\ast }\right) \right] .
\end{eqnarray*}%
The ensemble average of the wave intensity is then simply
\begin{equation}
\left\langle I\right\rangle _{c}(N,\beta ,\sigma )=N\psi ^{\ast }\left(
\beta ,\sigma \right) +o(N),  \label{M1_can_average_I}
\end{equation}%
meaning that the wave intensity is extensive when $\psi ^{\ast }>0$, and
non-extensive when $\psi ^{\ast }=0$. The mean intensity $I/N$ acts as an
order parameter (see Ref. \cite{phasetransition}): If $\beta _{c}<0$, then $%
\psi ^{\ast }$ is always positive; yet, if $\beta _{c}>0$, then for $\beta
<\beta _{c}$ (high temperature regime), $\psi ^{\ast }=0$ meaning that, as $%
t\rightarrow \infty $, the wave intensity $\psi $ should vanish and particle
motions should be ballistic, whereas for $\beta >\beta _{c}$ (low
temperature regime), $\psi ^{\ast }>0$ meaning that the wave intensity $\psi
$ should be finite as $t\rightarrow \infty $.

\section{The single wave case: saturation of the cold beam/plasma interaction%
}

\label{sec-coldbeam}

\subsection{Presentation}

The analysis of O'Neil and Drummond \textit{et al.} \cite{ONeil71,Drummond70}
have shown that, in the case of a monokinetic beam, only a single wave, the
wave having the largest growth rate, has a significant interaction with the
beam, at least up to the first trapping oscillations. We shall consider here
this single wave case and retain one mode ($M=1$) with rest frequency $%
\omega _{0}=\omega _{p}$ and wave number $k=1$. A simple linear analysis
\cite{ONeil71,these} determines then which initial beam streaming velocity $%
v_{b}$ leads to the maximal destabilization of this mode. This gives $%
v_{b}=\left( 1+2^{-4/3}\eta ^{1/3}\right) \omega _{p}$ and associated
maximal linear growth rate $\gamma _{L}=2^{-4/3}\sqrt{3}\eta ^{1/3}\omega
_{p}$. This derivation is analogous to considering the more realistic
scenario where the initial cold beam velocity is given and determining the
dominant mode, namely the mode, among a continuum of modes, that is the most
strongly destabilized. The expression of $v_{b}$ reveals the weakly resonant
nature of the linear instability: it concerns mostly particles having
velocities not equal to the initial wave velocity but about $\gamma_{L} /k $
away from it \cite{Zekri93}. Fig. \ref{fig_intensityCB} shows the time
evolution of the wave intensity starting from an infinitesimal value.
Nonlinear effects, namely trapping of particles in the wave trough, stop the
initial linear exponential growth. Trapping oscillations coincide with the
existence of a strong density inhomogeneity in the phase space. This is
visible on the phase space snapshot in Fig. \ref{fig_xvCB}.
\begin{figure}[tbh]
\begin{center}
\resizebox{9.0cm}{6cm} {\rotatebox{0}{\includegraphics[bb= 66 570
427 789]{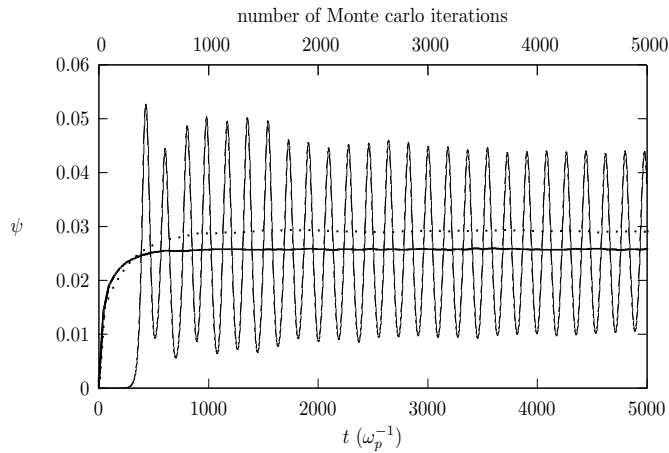}}} \label{fig_intensityCB}
\end{center}
\caption{Wave intensity $\protect\psi =I/N$ as a function of time (in $%
\protect\omega _{p}^{-1}$ units) in the cold beam case for $N=10000$ and $%
\protect\eta =3.85\times 10^{-5}$ (plain line). For the same parameters,
ensemble averages of $\protect\psi $ as a function of the number of Monte
Carlo iterations for the microcanonical ensemble (bold line) and for the
canonical ensemble with $U=E_{0}$ (dots) (See Sec. \protect\ref{sub_can_app}%
).\label{fig_intensityCB}}
\end{figure}
\begin{figure}[tbh]
\begin{center}
\resizebox{8.0cm}{5cm} {\rotatebox{0}{\includegraphics[bb= 53 570
427 780]{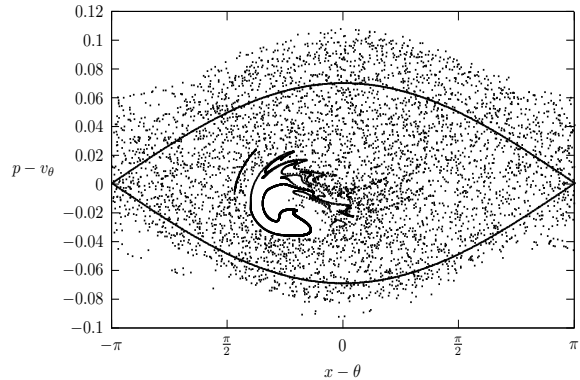}}}
\end{center}
\caption{Snapshot of the one-particle phase space in the wave frame for $%
t=5000\protect\omega _{p}^{-1}$. Instantaneous separatrices of the
wave resonance are shown. Dots represent the $N=10000$ particles,
which were initially distributed uniformly in $x$ with the same
velocity $v_{b}$.\label{fig_xvCB}}
\end{figure}

\subsection{Dynamical and Monte Carlo simulations; Canonical approach}

\label{sub_can_app}

We shall use the $M=1$ cold beam instability \cite{TMM} as a simple testbed
for the Monte Carlo code. Let us then first derive the canonical ensemble
average of the wave intensity in this case and look for an analytical
estimate as a function of the problem parameters and initial data.

The canonical ensemble average of energy is%
\begin{equation}
U=-\partial _{\beta }\ln Z_{c}=-\partial _{\beta }\ln Z_{0}(N,\beta
)-\partial _{\beta }\ln Z_{1}(N,\beta ,\sigma ),
\end{equation}%
with%
\begin{equation}
-\partial _{\beta }\ln Z_{0}(N,\beta )=\frac{N-1}{2\beta },
\label{contri_energie_Z0}
\end{equation}%
and%
\begin{equation}
-\partial _{\beta }\ln Z_{1}(N,\beta ,\sigma )=Ng(\beta ,\sigma ,\psi ^{\ast
}\mathbf{)}+o(N).  \label{contrib_energie_Z1}
\end{equation}%
This gives, with $\varepsilon \equiv U/N$,
\begin{equation}
\varepsilon =\frac{1}{2\beta }+\frac{1}{2}\left( \sigma -\psi ^{\ast
}\right) ^{2}+\omega _{p}\psi ^{\ast }-\beta ^{-1}\ln \left[ I_{0}\left(
\beta \omega _{0}^{3/2}\sqrt{2\eta \psi ^{\ast }}\right) \right] ,
\label{expression_energy}
\end{equation}%
that has to be solved together with (\ref{identity_phistar}). As given
above, we have $\sigma -\omega _{p}=c\eta ^{1/3}\omega _{p} $ with $%
c=2^{-4/3}$.

Let us look for solutions such that $\omega _{b\ast }^{2}\equiv \sqrt{2\eta
\psi ^{\ast }}\omega _{p}^{3/2}=a\eta ^{\alpha }\omega _{p}^{2}$. Then (\ref%
{identity_phistar}) yields
\begin{equation}
\psi ^{\ast }=c\eta ^{1/3}\omega _{p}+\frac{I_{1}\left( \beta a\eta ^{\alpha
}\omega _{p}^{2}\right) }{aI_{0}\left( \beta a\eta ^{\alpha }\omega
_{p}^{2}\right) }\eta ^{1-\alpha }\omega _{p}=\frac{a^{2}}{2}\eta ^{2\alpha
-1}\omega _{p}.  \label{scaling_phi_star}
\end{equation}%
A maximal ordering in (\ref{scaling_phi_star}) is obtained for $\beta a\eta
^{\alpha }\omega _{p}^{2}$ finite. It gives $1-\alpha =1/3=2\alpha -1$, that
is $\alpha =2/3$. It is important to note that this corresponds to the
trapping scaling for the wave intensity saturation, namely the asymptotic
trapping frequency, $\omega _{b\ast }$, scales as the linear growth rate: $%
\omega _{b\ast }\sim \eta ^{1/3}\omega _{p}\sim \gamma _{L}$. Introducing
the finite quantity $d\equiv \beta a\eta ^{2/3}\omega _{p}^{2}$, we have
then
\begin{equation}
\frac{I_{1}\left( d\right) }{I_{0}\left( d\right) }=a\left( \frac{a^{2}}{2}%
-c\right) \equiv F_{1}(d),  \label{relation_a_d_1}
\end{equation}%
and%
\begin{equation}
\frac{\varepsilon }{\omega _{p}^{2}}=\frac{1}{2}+c\eta ^{1/3}+\frac{a}{d}%
\left( \frac{1}{2}-\ln I_{0}\left( d\right) \right) \eta ^{2/3}+\frac{1}{2}%
\left( \frac{a^{2}}{2}-c\right) ^{2}\eta ^{2/3}.
\end{equation}%
Let us now assume that the canonical average of the energy density, $%
\varepsilon $, is close to its initial value, $\varepsilon _{0}=E_{0}/N$.
Because the wave intensity is initially vanishingly small, the
identification $\varepsilon =\varepsilon _{0}$ gives for the cold beam case
under consideration%
\begin{equation}
\frac{\left\langle p^{2}\right\rangle _{0}}{2}=\frac{\sigma ^{2}}{2}=\frac{%
\omega _{p}^{2}}{2}\left( 1+c\eta ^{1/3}\right) ^{2}.
\end{equation}%
This gives finally%
\begin{equation}
a\left( c-\frac{a^{2}}{4}\right) =\frac{1-2\ln I_{0}\left( d\right) }{d}%
\equiv F_{2}(d),  \label{relation_a_d_2}
\end{equation}%
to be solved together with (\ref{relation_a_d_1}). We get%
\begin{equation}
a=2^{2/3}\left[ F_{1}(d)+F_{2}(d)\right] ^{1/3}=c^{-1}\left(
F_{1}(d)+2F_{2}(d)\right) .
\end{equation}%
Using $c=2^{-4/3}$, this yields%
\begin{equation}
\frac{\left[ \frac{I_{1}\left( d\right) }{I_{0}\left( d\right) }+\left(
\frac{1-2\ln I_{0}\left( d\right) }{d}\right) \right] ^{1/3}}{\frac{%
I_{1}\left( d\right) }{I_{0}\left( d\right) }+2\left( \frac{1-2\ln
I_{0}\left( d\right) }{d}\right) }=2^{2/3}.
\end{equation}%
This gives numerically $d\simeq 1.56$ and $a\simeq 1.313$, yielding, with (%
\ref{scaling_phi_star}) and in the limit $N\rightarrow \infty $,
\begin{equation}
\psi ^{\ast }=\left\langle \frac{I}{N}\right\rangle _{c}\simeq 0.86\eta
^{1/3}\omega _{p}+o(1).  \label{cano_CB_appr}
\end{equation}%
Dynamical symplectic \cite{CaryDoxas93} simulations indicate that, after
some trapping periods, the average wave intensity $I/N$ scales approximately
as $0.78\eta ^{1/3}$ \cite{these}. For the parameter $\eta $ chosen in Fig. %
\ref{fig_intensityCB}, this gives $\left\langle I/N\right\rangle _{\mu
}\simeq 0.026$ as $N\rightarrow \infty $ which is indeed in agreement with
the time average of $\psi $. The microcanonical Monte Carlo results plotted
on the same Figure show an excellent agreement with this time average.
Canonical Monte Carlo results assuming $U=N\varepsilon _{0}$ is in fine
agreement with the analytical result (\ref{cano_CB_appr}). This is a
practical way to bypass the microcanonical or time-asymptotic evaluation of $%
\beta $ and relate directly canonical results to initial data. Here this
result shows a ten-per-cent discrepancy with the exact microcanonical
result. Moreover, we can equivalently express Eq. (\ref{cano_CB_appr}) in
terms of the ratio between the asymptotic trapping pulsation and the linear
growth rate yielding%
\begin{equation}
\frac{\omega _{b\ast }}{\gamma _{L}}=\frac{\left( 2\eta \psi ^{\ast }\right)
^{1/4}\omega _{p}^{3/4}}{2^{-4/3}\sqrt{3}\eta ^{1/3}\omega _{p}}\simeq \frac{%
\left( 2\times 0.86\right) ^{1/4}}{2^{-4/3}\sqrt{3}}=1.67.
\label{ratio_cano_cold_beam}
\end{equation}%
This is in agreement with the results of an early experiment by Mizuno and
Tanaka \cite{Mizuno72} designed to test the single wave model. A
monoenergetic beam is injected along a homogeneous magnetic field, making
the system unidimensional, into a plasma. In this electron gun, the
mechanism of the nonlinear saturation of a nearly single wave is reported
through the first measurement of the trapping of beam electrons. The
approximate experimental value, given in Ref. \cite{Mizuno72}, of the ratio $%
\omega _{b}/\gamma _{L}$ associated to the first trapping pulsation is 0.83.
As evidenced by Fig. \ref{fig_intensityCB}, this can here be used as a
fairly good approximation of the asymptotic value $\omega _{b\ast }/\gamma
_{L}$. Yet, proceeding to a direct measure from the data plotted on the
Figure 1 of Ref. \cite{Mizuno72}, we obtain a somewhat larger value for this
quantity, slightly above 1.2 and closer to the estimate (\ref%
{ratio_cano_cold_beam}).

Finally, it is interesting to note that an explicit low-dimensional modeling
of the dynamics, along the lines of Ref. \cite{TMM} stressing the phase
space division between the macroparticle and chaotic sea components, has
been recently proposed by Antoniazzi \textit{et al.} in the saturated regime
of a single pass free electron laser around perfect tuning \cite{AEFR2006}.
This model uses an Hamiltonian formulation that is a direct counterpart of
the present Hamiltonian single-wave particle framework.

\section{Landau damping of a single wave: Phase transition and O'Neil's
threshold}

\label{sec-ONeil-threshold}

In Ref. \cite{ONeil1965}, O'Neil considered the damping of a single wave
interacting effectively with a subset of resonant particles. He proposed a
qualitative dynamical threshold conditioning the time asymptotic behavior of
the wave: if the linear damping time $\left\vert \gamma _{L}\right\vert
^{-1} $ is far smaller than the initial nonlinear trapping time $\tau _{b0}$
of the particles in the wave trough, then the wave should not have enough
time to experience nonlinear effects and should be completely damped as $%
t\rightarrow \infty $. On the contrary, if the nonlinear time $\tau _{b0}$
is smaller than $\left\vert \gamma _{L}\right\vert ^{-1}$, the linear one,
nonlinear effects (trapping) should forever control the dynamics. This can
be already inferred from a linear analysis. As already apparent in the cold
beam case, the particles controlling the linear regime are those
quasi-resonant particles located at about $\gamma _{L}/k$ from $\omega
_{p}/k $ \cite{Zekri93}: if those particles are trapped in the initial wave
(with half resonance width $2\omega _{b0}$) right away, then the linear
regime should be bypassed. However, this result is qualitative. Actually, in
spite of major recent advances in the analytical and computational fields
\cite%
{Brodin97,Isi97,Manfredi97,LancellottiDorning,Brunetti2000,Valentini05,DeMarco2006,Ivanov06}%
, mostly in the full Vlasov-Poisson framework, as well as in experiments
\cite{Danielson04}, no definite answer presently exists to this
intrinsically nonlinear and time-asymptotic issue, in the sense that no
\textit{quantitative} threshold separating nonzero time-asymptotic wave
amplitude states from zero ones is known. This is basically due to the facts
that analytical reductions of the problem should e.g. make some restrictions
on the asymptotic behavior of the electric field whereas numerical and
experimental investigations should be obviously taken with great care when
addressing the $t\rightarrow \infty $ limit. For instance, the nice
experiments recently done in a trapped pure electron plasma \cite%
{Danielson04} diagnose the \textit{first} trapping oscillation frequency $%
\omega _{b}$ at the onset of the nonlinear stage (and more precisely the
dimensionless ratio $\gamma _{L}/\omega _{b}$) as a function of the initial
ratio $\gamma _{L}/\omega _{b0}$ rather than the, rigorously speaking,
\textit{time-asymptotic} state.

\begin{figure}[tbh]
\begin{center}
\resizebox{8.0cm}{5cm} {\rotatebox{0}{\includegraphics[bb= 258 600
558 802]{{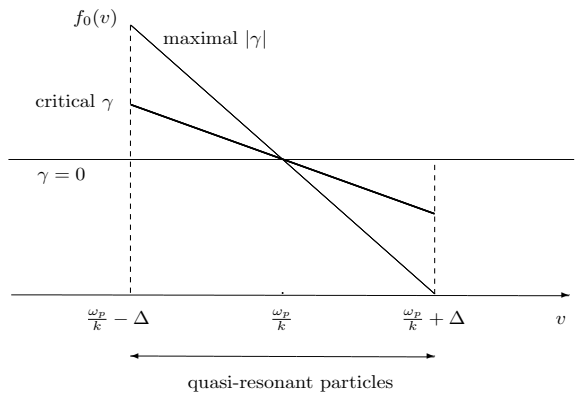}}}}
\end{center}
\caption{Family of tail distribution functions. We shall take $\Delta=4%
\protect\omega _{b0}/k$.\label{fig_ONEIL_setting}}
\end{figure}

We shall consider here the single wave-particle interaction model that
describes the effective (field envelope) dynamics of the Vlasov-Poisson
system in the case where only one wave exists. This is obviously an
idealized or academic situation, yet it is perfectly appropriate to discuss
O'Neil's picture. We shall show here that O'Neil's threshold can be related
to a second order phase transition for the single wave self-consistent
Hamiltonian model $H^{N,1}$, with order parameter the wave intensity $\psi $%
. This was already discussed in Ref. \cite{phasetransition}. We shall
however consider here specifically the O'Neil setting, that is we take a
given \textit{finite} initial wave intensity $\psi _{0}$ and a monotonously
decreasing, linear, tail particle distribution function $f_{0}(v)$, of which
we shall vary the slope and thus the linear Landau damping rate $\gamma _{L}$
(See Fig. \ref{fig_ONEIL_setting}). Because of locality properties in the
velocity space \cite{FirpoDoveil}, the wave has an effective interaction
only with particles that are not too far away from its separatrices. One can
reasonably think that particles having velocities more than, let's say, $%
4\omega _{b0}/k$ away from $\omega _{p}/k$ \cite{FirpoDoveil} should
evolve on Kolmogorov-Arnold-Moser (KAM) tori for the pendulum-like
1,5 degrees of freedom Hamiltonian (given by Eqs.
(\ref{eq_motion_1})-(\ref{eq_motion_2}) with $(I(t),\theta (t))$)
associated to $H^{N,1}$, materializing the separation between the
wave-particle interaction zone and the plasma bulk captured into the
collective variable $I$.

Let us then put $\omega _{p}=1$, define $\Delta =4\omega _{b0}=4\left( 2\eta
\psi _{0}\right) ^{1/4}$ and take $f_{0}(v)=av+b$ if $v\in \left[ 1-\Delta
;1+\Delta \right] $ and $f_{0}(v)=0$ otherwise, with $a<0$. Normalization
imposes $b=(1-2a\Delta )/(2\Delta )$ and positivity $-a<(2\Delta ^{2})^{-1}$%
. We have $\left\langle v\right\rangle _{0}=2a\Delta ^{3}/3+1$ \ and $%
T_{0}\equiv \left\langle v^{2}\right\rangle _{0}-\left\langle v\right\rangle
_{0}^{2}=-\Delta ^{2}\left( 4a^{2}\Delta ^{4}-3\right) /9$. Let us now write
the condition $\beta <\beta _{c}$ for which particle asymptotic behavior
should be ballistic. From (\ref{expression_energy}), we have in this regime
\begin{equation}
\left. \varepsilon \right\vert _{\psi ^{\ast }=0}=\frac{1}{2\beta }+\frac{%
\sigma ^{2}}{2}.  \label{eps_phi_star_0}
\end{equation}%
If we identify, once again, this canonical ensemble average with the initial
density of energy in the system $\varepsilon _{0}=$ $\left\langle
v^{2}\right\rangle _{0}/2+\psi _{0}$ and use $\sigma =\left\langle
v\right\rangle _{0}+$ $\psi _{0}$, we get the following condition on the
slope of the initial tail distribution:
\begin{equation}
\beta <\beta _{c}\Leftrightarrow Q(a)>\frac{27}{16}\eta \Delta ^{-9}
\label{beta_inf1}
\end{equation}%
where $Q$ is the third degree polynomial
\begin{eqnarray}
Q(x) &=&\left( x-c_{+}\right) \left( x-c_{-}\right) \left( x-c\right)
\label{beta_inf2} \\
\text{with \ }c_{\pm } &\equiv &c\pm \sqrt{3}/(2\Delta ^{2}),c\equiv
-3\Delta /(4^{5}\eta ).  \label{beta_inf3}
\end{eqnarray}%
Moreover, to ensure $\beta _{c}>0$, one should have $\sigma <\omega _{0}$
i.e. $\left\langle v\right\rangle _{0}+\psi _{0}<1$ so that
\begin{equation}
a<c=-3\Delta /(4^{5}\eta ).  \label{condTp}
\end{equation}%
$Q^{\prime }(x)$ vanishes for $d_{\pm }=c\pm 1/(2\Delta ^{2})$, the relative
maximum of $Q$ being reached in $d_{-}$. Taking into account the positivity
condition $-a<(2\Delta ^{2})^{-1}$, Eq. (\ref{condTp}) and $d_{-}<-(2\Delta
^{2})^{-1}$, one concludes that, in order to observe a phase transition for
this family of initial tail distribution functions\textit{, }it is necessary
and sufficient that $Q(-(2\Delta ^{2})^{-1})>27\eta \Delta ^{-9}/16$. The
necessary condition $Q\left( d_{-}\right) >\frac{27}{16}\eta \Delta ^{-9}$
is easier to formulate and reads
\begin{equation}
\Delta >3\times 2^{-2/3}\eta ^{1/3}\text{.}  \label{condnec}
\end{equation}%
The righthand side of (\ref{condnec}) is reminiscent of the scaling law of
the growth rate of the cold beam. This inequality is both a condition on the
beam width and initial wave amplitude. In order to observe a regime in which
the wave and the particles are decoupled in the sense that $I$ is no longer
extensive (ballistic regime instead of trapping regime), it is necessary to
exclude the case of a cold beam. In this latter case, even if the initial
wave amplitude is infinitesimal, it will not be completely damped: this
signals the singular character of the cold beam/wave interaction. Eq. (\ref%
{condnec}) means that a wave interacting resonantly with a cold beam cannot
be Landau damped to zero. Additionally, the condition (\ref{condnec}) thus
specifies what is meant by an "initially finite" wave amplitude. The
wave-particle interaction model \textit{\`{a} la O'Neil} presented here
rules out the classical research case of Landau damping in the limit of a
perturbative wave of infinitesimal amplitude. The present framework is
definitely non trivial since, if the wave is effectively damped
asymptotically to zero, nonlinear effects should place the study in the
frame of Isichenko's resolution \cite{Isi97} predicting a very slow,
asymptotically algebraic, damping.

Let us illustrate these results for values of $\Delta $ and $\eta $ allowing
a phase transition in the canonical approach just described.
\begin{figure}[tbh]
\begin{center}
\resizebox{8.0cm}{5cm} {\rotatebox{0}{\includegraphics[bb= 53 570
427 780]{{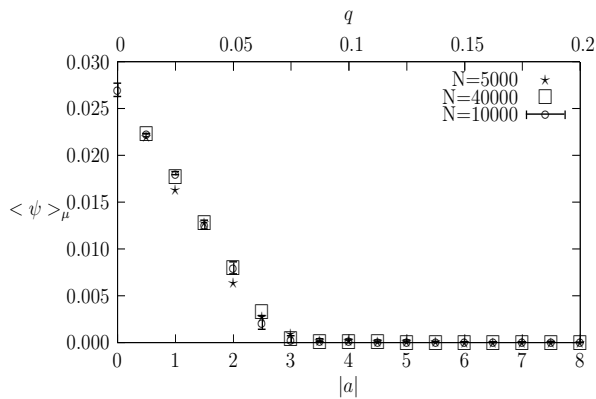}}}}
\end{center}
\caption{Monte Carlo microcanonical ensemble averages of the wave intensity $%
\protect\psi =I/N$ as a function of the slope $a$ of the initial
distribution function and corresponding $q$. For $N=10000$ particles, the
microcanonical ensemble averages of the fluctuations $<(\protect\psi -<%
\protect\psi >_{\protect\mu })^{2}>_{\protect\mu }^{1/2}$ are
indicated as errorbars.\label{fig_ONEIL}}
\end{figure}
\begin{figure}[tbh]
\begin{center}
\resizebox{8.0cm}{5cm} {\rotatebox{0}{\includegraphics[bb= 53 570
427 780]{{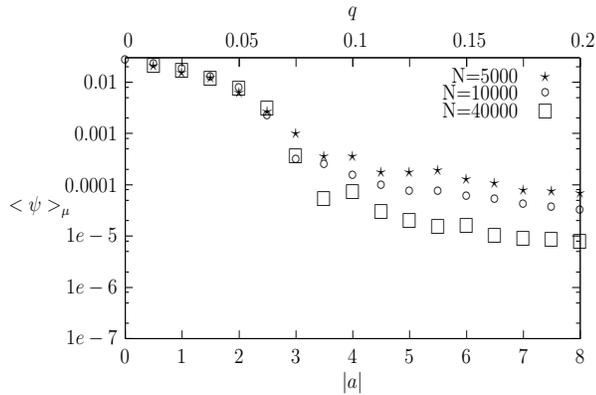}}}}
\end{center}
\caption{Same as Fig. \protect\ref{fig_ONEIL} in lin-log
scale.\label{fig_ONEIL_log}}
\end{figure}
Taking $\Delta =0.25$ and $\eta =0.001$ gives $Q(-(2\Delta
^{2})^{-1})=1011.5>\frac{27}{16}\eta \Delta ^{-9}=442.37$, which ensures the
existence of a phase transition. The tail distribution slope $a$ can vary
between $-8$ and $0$, and the critical slope associated with (\ref{beta_inf1}%
) is $a_{c}=-3.106$. This corresponds to a linear Landau damping rate $%
\gamma _{Lc}=\left( \pi a_{c}\eta \right) /2=-4.88\times 10^{-3}$. We are
now able to estimate the O'Neil threshold, namely the critical threshold of $%
q\equiv \left\vert \gamma _{L}\right\vert /\omega _{b0}$, as%
\begin{equation}
q_{c}=\frac{\left\vert \gamma _{Lc}\right\vert }{\omega _{b0}}\simeq 0.078.
\label{ON_threshold}
\end{equation}%
We can check that this value is very close to the threshold estimated in
Ref. \cite{phasetransition} in which the initial tail distribution (close
to, but not, linear), and thus the linear Landau damping rate, was \textit{%
given} whereas the initial wave amplitude was the \textit{control parameter}%
. This agreement may be interpreted as an indicator of the robustness of the
statistical mechanics approach. Figs. \ref{fig_ONEIL} and \ref{fig_ONEIL_log}
show the Monte Carlo microcanonical ensemble averages of the wave intensity
as a function of the slope of the initial tail distribution function and of
the parameter $q$. A\ nice agreement with the threshold (\ref{ON_threshold})
is found. The transition is emphasized by the logarithmic scale in Fig. \ref%
{fig_ONEIL_log}, as in the asymptotic ballistic regime, $\psi $ scales as $%
1/N$.

Using equilibrium statistical mechanics allows then to set in a quantitative
way O'Neil's threshold. However strong metastability effects and long
relaxation times towards equilibrium in the vicinity of the phase transition
\cite{phasetransition} temper the practical interest of this result,
especially in the purely collisionless Vlasovian $N\rightarrow \infty $
limit \cite{kinetic98} for which there is increasing evidence \cite%
{discrete,Bouchet2005,Yoon} that it may not commute with the $t\rightarrow
\infty $ limit. This explains that practical experimental and numerical
attempts to determine $q_{c}$ typically yield values that are about one
order larger than (\ref{ON_threshold}). It should be also remembered that,
as demonstrated by Isichenko's calculation \cite{Isi97}, the asymptotic
behavior of a \textit{damping} electric field is intrinsically very slow, so
very difficult to detect and discriminate practically from a
Bernstein-Greene-Kruskal (BGK) saturation stage \cite{BGK}. Equilibrium
statistical mechanics should prove however essential to address the large
time nonlinear fate of wave-particle interaction in the generic case of many
waves. Strong resonances overlap may then create large scale chaos inducing
an efficient sweeping of the phase space ensuring good ergodic properties
or, using the Lynden-Bell terminology, violent relaxation.

\section{Microcanonical predictions for a large number of waves: the
infinitesimal coupling limit}

\label{sec-small-coupling}

\subsection{Motivations and Monte Carlo simulations}

We shall consider here a regime for which the validity of quasilinear theory
has been questioned \cite{ALP79,DoxasCary97,Laval99}, when the nonlinear
timescale (resonance broadening time $\tau _{\mathrm{RB}}$) is much lower
than the linear one (of the order of the inverse of the linear growth rate $%
\gamma _{L}$), that is $\mu \equiv (\tau _{\mathrm{RB}}\gamma _{L})^{-1}\gg
1 $. Equilibrium predictions are expected to be particularly relevant for
this problem since chaotic mixing and system ergodization should prove
effective typically on the nonlinear timescale.

Let us begin by presenting microcanonical Monte Carlo simulations done for
parameters very close to those used by Doxas and Cary in Ref. \cite%
{DoxasCary97}. Specifically, the wave spectrum is composed of $M=584$ waves,
that are initially strongly overlapping with a Chirikov overlap parameter of
about 500, with wavenumbers between 2.78 and 3.70. The strong overlap is
needed to model a continuous spectrum. The beam distribution in velocity is
the same as in Ref. \cite{DoxasCary97}, chosen to give a constant linear
growth rate across the spectrum: $f_{0}(v)=C_{1}+C_{2}\left(
1-v_{a}/v\right) $ if $v\in \left[ v_{a};v_{b}\right] $ and $f_{0}(v)=0$
otherwise, with $C_{1}$ and $C_{2}$ fixed by normalization and the value of
the growth rate. For the Monte Carlo code, these particular initial
conditions just fix the two microcanonical constants of motion, namely the
total energy and momentum. Then, as detailed in the analysis of Ref. \cite%
{DoxasCary97}, the large-$\mu $ regime is entered as the coupling (control)
parameter $\eta $ becomes vanishingly small. We are thus investigating here
the small coupling regime.
\begin{figure}[tbh]
\begin{center}
\resizebox{8.5cmn}{5cm} {\rotatebox{0}{%
\includegraphics[90,2][530,275]{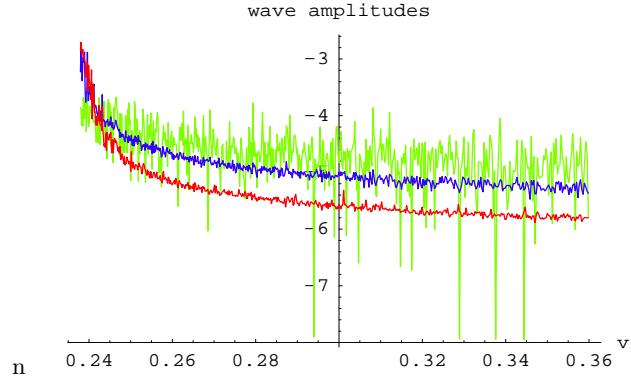}}}
\end{center}
\caption{(Color Online) Wave spectrum obtained through
microcanonical Monte Carlo simulations, for $\protect\eta =2.6\times
10^{-4}$ and initial conditions described in the text, for two
values of the number $N$ of particles: with $N=5000$ in the
transient regime (after only a few iterations) (upper curve), with
$N=5000$ at equilibrium (middle curve) and for $N=20000$ at
equilibrium (lower curve). Wave amplitudes are given in logarithmic
scale.\label{fig_wave_spectrum}}
\end{figure}
Results may be summarized as follows: in the transient stage of the Monte
Carlo runs, that may be compared with some care to the real time dynamics,
one observes "noisy" wave spectra that may be related to the spontaneous
spectra discretization reported in previous studies \cite{DoxasCary97}. In
the thermodynamical framework, this appears only as a transient stage
towards equilibration. The major point depicted by Fig. \ref%
{fig_wave_spectrum} is that the energy apparently tends to concentrate at
equilibrium into a very small number of small wavelength modes while the
long wavelength wave intensities are no longer extensive.

\subsection{Exact microcanonical derivation}

Let us consider analytically the small coupling limit, namely $\eta
\rightarrow 0^{+}$. This is a singular limit that amounts to neglect (at
zero order in $\eta $) the wave-particle interaction potential in the
self-consistent Hamiltonian (\ref{hamiltonian}) while retaining the total
wave-particle momentum conservation constraint. In the microcanonical
ensemble, the partition function reads, with $\omega _{0j}=\omega _{p}$ for
any wave number $j$,%
\begin{equation}
Z_{\mu }(E,P)=\int \delta \left( E-\sum\limits_{l=1}^{N}\frac{p_{l}^{2}}{2}%
-\omega _{p}\sum_{j=1}^{M}I_{j}\right) \delta \left(
P-\sum\limits_{l=1}^{N}p_{l}-\sum_{j=1}^{M}k_{j}I_{j}\right) d^{N}\mathbf{p}%
d^{M}\mathbf{I.}  \label{defin_Z_mu_eta_0}
\end{equation}%
Let us introduce $I_{j}^{\prime }=I_{j}$ for $1\leq j\leq M-1$ and $%
I_{M}^{\prime }$ such that $\sum_{j=1}^{M}I_{j}=\sum_{j=1}^{M-1}I_{j}^{%
\prime }+I_{M}=MI_{M}^{\prime }$. We get%
\begin{equation}
Z_{\mu }(E,P)=M\int \delta \left( E-\sum\limits_{l=1}^{N}\frac{p_{l}^{2}}{2}%
-\omega _{p}MI_{M}^{\prime }\right) \delta \left(
P-\sum\limits_{l=1}^{N}p_{l}-\sum_{j=1}^{M-1}\left( k_{j}-k_{M}\right)
I_{j}^{\prime }-k_{M}MI_{M}^{\prime }\right) d^{N}\mathbf{p}d^{M}\mathbf{I}%
^{\prime }\mathbf{.}
\end{equation}%
Introducing%
\begin{equation}
B\equiv \frac{k_{M}}{\omega _{p}}E+\frac{N\omega _{p}}{2k_{M}}-P,
\label{def_B}
\end{equation}%
integration over $I_{M}^{\prime }$ gives%
\begin{eqnarray}
Z_{\mu }(E,P) &=&M\int \delta \left(
P-\sum\limits_{l=1}^{N}p_{l}-\sum_{j=1}^{M-1}\left( k_{j}-k_{M}\right)
I_{j}^{\prime }-\frac{k_{M}}{\omega _{p}}E+\frac{k_{M}}{\omega _{p}}%
\sum\limits_{l=1}^{N}\frac{p_{l}^{2}}{2}\right) d^{N}\mathbf{p}%
\prod\limits_{j=1}^{M-1}dI_{j}^{\prime }  \notag \\
&=&M\int \delta \left( \sum_{j=1}^{M-1}\left( k_{M}-k_{j}\right)
I_{j}^{\prime }-B+\sum\limits_{l=1}^{N}\left( \sqrt{\frac{k_{M}}{2\omega _{p}%
}}p_{l}-\sqrt{\frac{\omega _{p}}{2k_{M}}}\right) ^{2}\right) d^{N}\mathbf{p}%
d^{M-1}\mathbf{I}^{\prime }  \notag \\
&=&M\left( \frac{2\omega _{p}}{k_{M}}\right) ^{N/2}\int \delta \left(
\sum_{j=1}^{M-1}\left( k_{M}-k_{j}\right) I_{j}^{\prime
}-B+\sum\limits_{l=1}^{N}\tilde{p}_{l}^{2}\right) d^{N}\mathbf{\tilde{p}}%
d^{M-1}\mathbf{I}^{\prime }  \label{inter_Z_microcanonique}
\end{eqnarray}%
It is requested that
\begin{equation}
0\leq \sum_{j=1}^{M-1}\left( k_{M}-k_{j}\right) I_{j}^{\prime }\leq B,
\end{equation}%
so that, in particular, with $\omega _{p}/k_{M}=v_{\theta \min }$,
\begin{equation}
v_{\theta \min }\sigma -\frac{v_{\theta \min }^{2}}{2}\leq \varepsilon .
\label{ineq_eta_zero}
\end{equation}%
This inequality (\ref{ineq_eta_zero}) is basically satisfied when the
particle impulsions are larger than the smallest wave velocity, which is the
natural setting of the wave-particle interaction model. Eq. (\ref%
{inter_Z_microcanonique}) is easily integrated over the $\tilde{p}_{l}$'s
yielding, assuming $N$ even,
\begin{equation}
Z_{\mu }(E,P)=\frac{2M}{\pi \left( N/2-1\right) !}\left( \frac{2\pi \omega
_{p}}{k_{M}}\right) ^{N/2}\int f\left( \sum_{j=1}^{M-1}\left(
k_{M}-k_{j}\right) I_{j}^{\prime }\right) d^{M-1}\mathbf{I}^{\prime },
\end{equation}%
with $f(x)\equiv \left( B-x\right) ^{(N-1)/2}$ and where the integration
boundaries are given $0\leq \sum_{j=1}^{M-1}\left( k_{M}-k_{j}\right)
I_{j}^{\prime }\leq B$ with $0\leq I_{j}^{\prime }$, $\forall j$, $1\leq
j\leq M-1$. We begin by making a change a variable, putting $\left(
k_{M}-k_{j}\right) I_{j}^{\prime }=\tilde{I}_{j}$. This gives%
\begin{equation}
Z_{\mu }(E,P)=\frac{2M}{\pi \left( N/2-1\right) !}\left( \frac{2\pi \omega
_{p}}{k_{M}}\right) ^{N/2}\prod\limits_{j=1}^{M-1}\left( k_{M}-k_{j}\right)
^{-1}\int_{0\leq \sum_{j=1}^{M-1}\tilde{I}_{j}\leq B}f\left( \sum_{j=1}^{M-1}%
\tilde{I}_{j}\right) d^{M-1}\mathbf{\tilde{I}.}
\end{equation}%
We have%
\begin{eqnarray*}
&&\int_{0\leq \sum_{j=1}^{M-1}\tilde{I}_{j}\leq B}f\left( \sum_{j=1}^{M-1}%
\tilde{I}_{j}\right) d^{M-1}\mathbf{\tilde{I}} \\
&=&\int d\tilde{I}_{M-1}\int \ldots \int\limits_{0}^{B-\sum_{j=2}^{M-1}%
\tilde{I}_{j}}d\tilde{I}_{1}\left( B-\sum_{j=1}^{M-1}\tilde{I}_{j}\right)
^{(N-1)/2} \\
&=&\frac{1}{(N-1)/2+1}\int d\tilde{I}_{M-1}\int \ldots
\int\limits_{0}^{B-\sum_{j=3}^{M-1}\tilde{I}_{j}}d\tilde{I}_{2}\left(
B-\sum_{j=2}^{M-1}\tilde{I}_{j}\right) ^{(N-1)/2+1} \\
&=&\prod_{j=1}^{M-2}\frac{1}{(N-1)/2+j}\int_{0}^{B}d\tilde{I}_{M-1}\left( B-%
\tilde{I}_{M-1}\right) ^{(N-1)/2+M-2} \\
&=&\prod_{j=1}^{M-1}\frac{1}{(N-1)/2+j}B^{(N-1)/2+M-1}.
\end{eqnarray*}%
Finally, we get%
\begin{equation}
Z_{\mu }(E,P)=\frac{2M}{\pi \left( N/2-1\right) !}\left( \frac{2\pi \omega
_{p}}{k_{M}}\right) ^{N/2}\times \prod_{j=1}^{M-1}\frac{\left(
k_{M}-k_{j}\right) ^{-1}}{(N-1)/2+j}\times \left[ \frac{k_{M}}{\omega _{p}}E+%
\frac{N\omega _{p}}{2k_{M}}-P\right] ^{(N-1)/2+M-1}.
\label{partition_micro_eta0+}
\end{equation}%
Following the calculations presented in the Appendix \ref{app-micro}, we
derive the wave intensities for any $j_{0}$ strictly smaller than $M$ as%
\begin{equation}
\left\langle \psi _{j_{0}}\right\rangle _{\mu }^{(N)}\sim _{N\gg 1}\frac{1}{N%
}\frac{T_{\mu }}{\omega _{p}}\frac{v_{\theta j_{0}}}{v_{\theta
j_{0}}-v_{\theta \min }}.  \label{micro_average_I_j0}
\end{equation}%
\begin{figure}[tbh]
\begin{center}
\resizebox{8.5cmn}{5cm} {{\includegraphics[90,4][403,196]{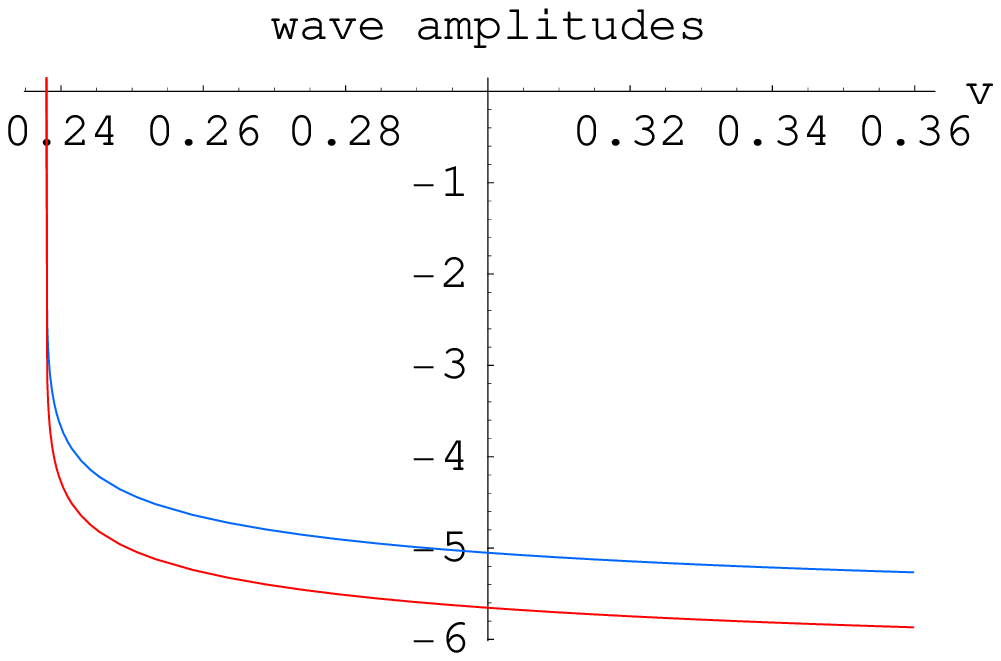}}} %
\end{center}
\caption{(Color Online) Equilibrium wave spectrum as derived in Eq. (\protect
\ref{micro_average_I_j0}) for the same initial wave and particle
distributions as in Fig. \protect\ref{fig_wave_spectrum}, for two values of
the number $N$ of particles: for $N=5000$ (upper curve) and for $%
N=20000 $ (lower curve). Wave amplitudes are given in logarithmic
scale. Numerical data for the initial conditions are
$v_{\protect\theta \min }=0.2379$, $\protect\varepsilon =0.0555$ and
$\protect\sigma =0.333$.\label{fig_wave_spectrum_theory}}
\end{figure}
As shown on Fig. \ref{fig_wave_spectrum_theory}, this result nicely agrees
with Monte Carlo microcanonical expectations in the limit $\eta \rightarrow
0^{+}$. We have introduced the microcanonical temperature $T_{\mu }$ through
\footnote{%
We always assume $N\rightarrow \infty $, keeping $M$ large but finite, and
take $k_{B}=1$.}
\begin{equation}
\frac{1}{T_{\mu }}=\frac{\partial \ln Z_{\mu }}{\partial E}\sim \frac{1}{2}%
\frac{\frac{k_{M}}{\omega _{p}}}{\frac{k_{M}}{\omega _{p}}\varepsilon +\frac{%
\omega _{p}}{2k_{M}}-\sigma },
\end{equation}%
so that%
\begin{equation}
\frac{T_{\mu }}{2}\sim \varepsilon +\frac{v_{\theta \min }^{2}}{2}-\sigma
v_{\theta \min }.  \label{temp_micro}
\end{equation}%
This clarifies the inequality (\ref{ineq_eta_zero}), that just means that
the temperature $T_{\mu }$ should be positive. Let us now complete this
study by moving to a grand canonical ensemble, that is both canonical with
respect to the energy and the total momentum.

\subsection{Grand canonical approach}

We define
\begin{equation}
Z_{G}\left( \beta ,\lambda \right) \equiv \int \exp \left( -\beta H-\lambda
P\right) d^{N}\mathbf{p}d^{M}\mathbf{I,}  \label{defi_full_cano}
\end{equation}%
with $H=\sum_{l=1}^{N}\frac{p_{l}^{2}}{2}+\omega _{p}\sum_{j=1}^{M}I_{j}$
and $P=\sum_{l=1}^{N}p_{l}+\sum_{j=1}^{M}k_{j}I_{j}$. This gives immediately%
\begin{equation}
Z_{G}\left( \beta ,\lambda \right) =\left( \frac{2\pi }{\beta }\right)
^{N/2}\exp \left( \frac{N\lambda ^{2}}{2\beta }\right) \prod_{j=1}^{M}\left(
\beta \omega _{p}+\lambda k_{j}\right) ^{-1}.  \label{exp_full_cano}
\end{equation}%
In this ensemble, we easily derive
\begin{equation}
\left\langle I_{j_{0}}\right\rangle _{G}=\frac{1/\beta }{\left( \omega
_{p}+k_{j_{0}}\lambda /\beta \right) }=\frac{v_{\theta j_{0}}/\beta }{\omega
_{p}\left( v_{\theta j_{0}}+\lambda /\beta \right) },  \label{inten_grandcan}
\end{equation}%
with (see Appendix \ref{app-cano-general})
\begin{equation}
\frac{\lambda }{\beta }=-\left\langle \frac{1}{N}\sum_{l=1}^{N}p_{l}\right%
\rangle _{G},  \label{p_equilibre_grandcan}
\end{equation}%
and the ensemble average of the reduced kinetic energy is found to be%
\begin{eqnarray*}
\left\langle K\right\rangle _{G} &\equiv &\left\langle \sum_{l=1}^{N}\frac{1%
}{2}\left( p_{l}-\frac{1}{N}\sum_{k=1}^{N}p_{k}\right) ^{2}\right\rangle _{G}
\\
&=&\left\langle \sum_{l=1}^{N}\frac{p_{l}^{2}}{2}\right\rangle _{G}-\frac{1}{%
2N}\left\langle \left( \sum_{l=1}^{N}p_{l}\right) ^{2}\right\rangle _{G}\sim
\frac{N}{\beta }.
\end{eqnarray*}

\subsection{Discussion and conclusion}

Considering a spectrum of waves, we have focused on the small coupling limit
as a benchmark for the statistical mechanics of weak Langmuir turbulence. A
trivial correspondence, assuming the equivalence of ensembles, between Eqs. (%
\ref{micro_average_I_j0}) and (\ref{inten_grandcan}), using (\ref%
{p_equilibre_grandcan}), shows that electrons should, as time proceeds,
escape from the original resonances towards lower speeds. Meanwhile, the
wave spectrum collapses towards short-wavelengths and the wave energy
eventually concentrates into the mode of minimal phase velocity, that
basically matches plasma thermal velocity. Consistently, particles
(electrons) behaving almost freely loose their resonant character and follow
the decay of waves towards small velocities.

Let us then briefly comment on the relation of these results with the
investigation on the limits of quasilinear theory undertaken by Doxas and
Cary \cite{DoxasCary97}. It is obviously uneasy to draw conclusions on the
out-of-equilibrium behavior, such as diffusion, from equilibrium statistical
mechanics results. Some statements are still possible: When the nonlinear
timescale is far larger than the linear one, quasilinear theory is expected
not to apply as the timescale of validity of (almost) zero-order effects,
such as the reaction of the resonant wave spectrum on the zero-order
distribution function, becomes negligible. Actually, we infer from the
equilibrium microcanonical results that, within some nonlinear
thermalization timescale, the initial particle distribution function is
substantially modified whereas the wave spectrum collapses towards the short
wavelengths. The thermalization stage is expected to exhibit a noisy wave
spectrum, as in Monte Carlo transients, as the wave spectrum collapse
proceeds through higher order wave-wave couplings, which may explain the
dynamical findings of diffusion enhancements \cite{DoxasCary97}.

Now, we can consider the small coupling results from a more general
perspective. Let us consider an initial very weak beam-plasma or
bump-on-tail instability destabilizing some spectrum of waves towards
suprathermal levels in a wave-particle interaction dynamics captured by the
self-consistent Hamiltonian (\ref{hamiltonian}). Then, the previous
calculation predicts that this out-of-equilibrium dynamics cannot sustain
itself as the wave energy should drop and cascade towards short-wavelengths
with phase velocities reaching the plasma bulk thermal velocity. Obviously,
the self-consistent Hamiltonian model reaches at this point its validity
limit. This behavior is an effect of considering a spectrum of modes rather
than single modes, as in Secs. \ref{sec-coldbeam} and \ref%
{sec-ONeil-threshold}, for which trapping ensures the long-standing
stabilization of the instability.

Recently a pending debate emerged concerning the time-asymptotic state of a
large amplitude Landau damped wave \cite{Danielson04}. It is commonly
expected that the system will eventually enter some stationary BGK
equilibrium with a time-asymptotic finite wave amplitude provided the
initial amplitude is large enough. This agrees with the statistical
mechanics predictions for a single wave derived in Sec. \ref%
{sec-ONeil-threshold} under the O'Neil setting. The existence of BGK steady
states is supported by some theoretical analysis and numerical results \cite%
{Manfredi97,LancellottiDorning,Brunetti2000}. However, Brodin's
numerical simulations \cite{Brodin97} suggested that the wave
amplitude never settles to a steady value as the energy of the
resonant particles diffuses slowly into higher harmonics. Moreover
Isichenko \cite{Isi97} argued that nonlinearities should not stop
the damping of the electric field towards a vanishing
time-asymptotic amplitude. Our point of view here is that, within
the weak coupling hypothesis - that besides fully justifies the
linear response of the background plasma -, the Landau damping of a
\textit{spectrum of waves} is predicted \textit{not to reach} a BGK
finite amplitude steady state with the resonant tail electrons
drifting with wave energy towards plasma bulk.

\appendix

\section{Exact microcanonical ensemble averages of the wave intensities in
the limit $\protect\eta \rightarrow 0^{+}$}

\label{app-micro}

We wish to calculate
\begin{equation}
\left[ I_{j_{0}}^{\prime }\right] ^{(N)}\equiv \frac{2M}{\pi \left(
N/2-1\right) !}\left( \frac{2\pi \omega _{p}}{k_{M}}\right) ^{N/2}\int
I_{j_{0}}^{\prime }\left( B-\sum_{j=1}^{M-1}\left( k_{M}-k_{j}\right)
I_{j}^{\prime }\right) ^{(N-1)/2}d^{M-1}\mathbf{I}^{\prime }.
\end{equation}%
We have%
\begin{equation}
\frac{\partial Z_{\mu }^{(N+2)}}{\partial k_{j_{0}}}=\frac{2M}{\pi \left(
N/2+1\right) !}\left( \frac{2\pi \omega _{p}}{k_{M}}\right) ^{N/2+1}\frac{N+1%
}{2}\int I_{j_{0}}^{\prime }\left( B-\sum_{j=1}^{M-1}\left(
k_{M}-k_{j}\right) I_{j}^{\prime }\right) ^{(N-1)/2}d^{M-1}\mathbf{I}%
^{\prime },
\end{equation}%
so that%
\begin{equation}
\left[ I_{j_{0}}^{\prime }\right] ^{(N)}=\left( \frac{N}{2}+1\right) \frac{N%
}{2}\frac{k_{M}}{\pi \omega _{p}\left( N+1\right) }\frac{\partial Z_{\mu
}^{(N+2)}(E,P)}{\partial k_{j0}},
\end{equation}%
and the microcanonical ensemble average of the $j_{0}^{th}$ wave intensity
is
\begin{equation}
\left\langle I_{j_{0}}^{\prime }\right\rangle _{\mu }^{(N)}\equiv \frac{%
\left[ I_{j_{0}}^{\prime }\right] ^{(N)}}{Z_{\mu }^{(N)}}=\frac{N/2}{\left(
\frac{2\pi \omega _{p}}{k_{M}}\right) \left( \frac{N+1}{2}\right) }\frac{%
\partial Z_{\mu }^{(N+2)}/\partial k_{j_{0}}}{Z_{\mu }^{(N)}}.
\end{equation}%
We have, from Eq. (\ref{partition_micro_eta0+}),%
\begin{eqnarray*}
\partial Z_{\mu }^{(N+2)}/\partial k_{j_{0}} &=&\frac{2M}{\pi \left(
N/2+1\right) !}\left( \frac{2\pi \omega _{p}}{k_{M}}\right) ^{N/2+1} \\
&&\times \prod_{j=1}^{M-1}\frac{\left( k_{M}-k_{j}\right) ^{-1}}{(N+1)/2+j}%
\times \left[ \frac{k_{M}}{\omega _{p}}E+\frac{N\omega _{p}}{2k_{M}}-P\right]
^{(N+1)/2+M-1}\left( k_{M}-k_{j_{0}}\right) ^{-1}
\end{eqnarray*}%
so that%
\begin{equation}
\left\langle I_{j_{0}}^{\prime }\right\rangle _{\mu }^{(N)}=\frac{2N}{\left(
N+1\right) }\times \frac{(N-1)/2+1}{(N-1)/2+M}\times \left[ \frac{k_{M}}{%
\omega _{p}}\varepsilon +\frac{\omega _{p}}{2k_{M}}-\sigma \right] \times
\left( k_{M}-k_{j_{0}}\right) ^{-1}.
\end{equation}%
This gives finally, for $j_{0}\neq M$,%
\begin{eqnarray*}
\left\langle \omega _{p}I_{j_{0}}^{\prime }\right\rangle _{\mu }^{(N)} &\sim
&_{N\gg 1}\frac{k_{M}}{k_{M}-k_{j_{0}}}T_{\mu } \\
&=&\frac{v_{\theta j_{0}}}{v_{\theta j_{0}}-v_{\theta \min }}T_{\mu },
\end{eqnarray*}%
with $T_{\mu }$ the microcanonical temperature given in (\ref{temp_micro}).

\section{Grand canonical approach: explicit results}

\label{app-cano-general}

Let us consider%
\begin{eqnarray*}
Z_{G}\left( \beta ,\lambda ,a_{1},a_{2},a_{3},a_{4}\right) &=&\int \exp
\left( -\beta a_{1}\sum_{l=1}^{N}\frac{p_{l}^{2}}{2}-\beta a_{2}\omega
_{p}\sum_{j=1}^{M}I_{j}-\lambda a_{3}\sum_{l=1}^{N}p_{l}-\lambda
a_{4}\sum_{j=1}^{M}k_{j}I_{j}\right) d^{N}\mathbf{p}d^{M}\mathbf{I} \\
&=&\left[ \int_{-\infty }^{\infty }\exp \left( -\beta a_{1}\frac{p^{2}}{2}%
-\lambda a_{3}p\right) dp\right] ^{N}\prod_{j=1}^{M}\int_{0}^{\infty }\exp
\left( -\beta a_{2}\omega _{p}I_{j}-\lambda a_{4}k_{j}I_{j}\right) dI_{j} \\
&=&\left( \frac{\sqrt{2\pi }}{\beta a_{1}}\right) ^{N}\exp \left( \frac{%
N\lambda ^{2}a_{3}^{2}}{2\beta a_{1}}\right) \prod_{j=1}^{M}\left( \beta
a_{2}\omega _{p}+\lambda a_{4}k_{j}\right) ^{-1}.
\end{eqnarray*}%
We have then immediately
\begin{equation}
\left\langle \sum_{l=1}^{N}p_{l}\right\rangle _{G}=\left. -\frac{1}{\lambda }%
\frac{\partial \ln Z_{G}}{\partial a_{3}}\right\vert _{a_{1,2,3,4}=1}=-\frac{%
N\lambda }{\beta },
\end{equation}%
\begin{equation}
\left\langle \left( \sum_{l=1}^{N}p_{l}\right) ^{2}\right\rangle _{G}=\left.
\frac{1}{\lambda ^{2}}\frac{\frac{\partial ^{2}Z_{G}}{\partial a_{3}^{2}}}{%
\tilde{Z}_{c}}\right\vert _{a_{1,2,3,4}=1}=\frac{N^{2}\lambda ^{2}}{\beta
^{2}}+\frac{N}{\beta },
\end{equation}%
and%
\begin{equation}
\left\langle \sum_{l=1}^{N}\frac{p_{l}^{2}}{2}\right\rangle _{G}=\left. -%
\frac{1}{\beta }\frac{\partial \ln Z_{G}}{\partial a_{1}}\right\vert
_{a_{1,2,3,4}=1}=\frac{N}{\beta }+\frac{N}{\beta }\frac{\lambda ^{2}}{2\beta
}.
\end{equation}%
This gives%
\begin{equation}
\left\langle K\right\rangle _{G}=\left\langle \sum_{l=1}^{N}\frac{p_{l}^{2}}{%
2}\right\rangle _{G}-\frac{1}{2N}\left\langle \left(
\sum_{l=1}^{N}p_{l}\right) ^{2}\right\rangle _{G}=\frac{N}{\beta }-\frac{1}{%
2\beta }\sim \frac{N}{\beta }.
\end{equation}

\acknowledgments

MCF thanks Y. Elskens for introducing her to the wave-particle
self-consistent framework and for related fruitful discussions. GA wishes to
thank the Laboratoire de Physique et de Technologie des Plasmas for hosting
him.


\begin{thebibliography}{99}
\bibitem{Sagdeev} R. Z. Sagdeev and A. A. Galeev, Nonlinear Plasma
Theory (Benjamin, New York, 1969).

\bibitem{Mynick78} H. E. Mynick and A. N. Kaufman, Phys. Fluids \textbf{21},
653 (1978).

\bibitem{CaryDoxas93} J. R. Cary and I. Doxas, J. Comput. Phys. \textbf{107}%
, 98 (1993).

\bibitem{TMM} J. L. Tennyson, J. D. Meiss and P. J. Morrison, Physica D
\textbf{71}, 1 (1994).

\bibitem{AEE98} M. Antoni, Y. Elskens and D. F. Escande, Phys. Plasmas
\textbf{5}, 841 (1998).

\bibitem{book} Y. Elskens and D. F. Escande, Microscopic Dynamics of
Plasmas and Chaos (IOP, Bristol, 2002).


\bibitem{discrete} F. Doveil, M.-C. Firpo, Y. Elskens, D. Guyomarc'h, M.
Poleni and P. Bertrand, Phys. Lett. A \textbf{284}, 279 (2001); M.-C. Firpo,
F. Doveil, Y. Elskens, P. Bertrand, M. Poleni and D. Guyomarc'h, Phys. Rev.
E \textbf{64}, 026407 (2001); M.-C. Firpo and Y. Elskens, Transport Theory
and Stat. Phys. \textbf{32}, 399 (2003).

\bibitem{phasetransition} M.-C. Firpo and Y. Elskens, Phys. Rev. Lett.
\textbf{84}, 3318 (2000).

\bibitem{KRAFFT05} C. Krafft, A. Volokitin, and A. Zaslavsky, Phys. Plasmas \textbf{12}, 112309 (2005).

\bibitem{ONeil1965} T. M. O'Neil, Phys. Fluids \textbf{8}, 2255 (1965).

\bibitem{kinetic98} M.-C. Firpo and Y. Elskens, J. Stat. Phys. \textbf{93},
193 (1998).

\bibitem{ONeil71} {T. M. O'Neil, J. H. Winfrey, and J. H. Malmberg, Phys.
Fluids \textbf{14}, 1204 (1971).}

\bibitem{Drummond70} W. E. Drummond, J. H. Malmberg, T. M. O'Neil, and J. R.
Thompson, Phys. Fluids \textbf{13}, 2422 (1970).

\bibitem{delCastillo2000} D. del-Castillo-Negrete, Physica A \textbf{280},
10 (2000).

\bibitem{these} M.-C. Firpo, \textit{Etude dynamique et statistique de
l'interaction onde-particule}, Th\`{e}se de doctorat de l'universit\'{e} de
Provence (Marseille, 1999).

\bibitem{Zekri93} {S. Zekri, \textit{Approche hamiltonienne de la turbulence
faible de Langmuir}, Th\`{e}se de doctorat de l'universit\'{e} de Provence
(Marseille, 1993).}

\bibitem{Mizuno72} K. Mizuno and S. Tanaka, Phys. Rev. Lett. \textbf{29}, 45
(1972).

\bibitem{AEFR2006} A. Antoniazzi, Y. Elskens, D. Fanelli, and S. Ruffo, Eur.
J. Phys. B \textbf{50}, 603 (2006).

\bibitem{Brodin97} G. Brodin, Phys. Rev. Lett. \textbf{78}, 1263 (1997).

\bibitem{Isi97} M. B. Isichenko, Phys. Rev. Lett. \textbf{78}, 2369 (1997).

\bibitem{Manfredi97} G. Manfredi, Phys. Rev. Lett. \textbf{79}, 2815 (1997).

\bibitem{LancellottiDorning} C. Lancellotti and J. J. Dorning, Phys. Rev.
Lett. \textbf{81}, 5137 (1998); Phys. Rev. E \textbf{68}, 026406 (2003).

\bibitem{Brunetti2000} M. Brunetti, F. Califano, and F. Pegoraro, Phys. Rev.
E \textbf{62}, 4109 (2000).

\bibitem{Valentini05} F. Valentini, V. Carbone, P. Veltri, and A. Mangeney, Phys. Rev. E \textbf{71}, 017402
(2005).

\bibitem{DeMarco2006} R. De Marco, V. Carbone, and P. Veltri, Phys. Rev.
Lett. \textbf{96}, 125003 (2006).

\bibitem{Ivanov06} A. V. Ivanov and I. H. Cairns, Phys. Rev. Lett. \textbf{96}, 175001 (2006).

\bibitem{Danielson04} J. R. Danielson, F. Anderegg, and C. F. Driscoll,
Phys. Rev. Lett. \textbf{92}, 245003 (2004).

\bibitem{Bouchet2005} F. Bouchet and T. Dauxois, Phys. Rev. E \textbf{72},
045103(R) (2005).

\bibitem{Yoon} P. H. Yoon, T. Rhee, and C.-M. Ryu, Phys. Plasmas \textbf{12}, 062310 (2005); Phys. Rev. Lett. \textbf{%
95}, 215003 (2005).

\bibitem{FirpoDoveil} M.-C. Firpo and F. Doveil, Phys. Rev. E \textbf{65},
016411 (2002).

\bibitem{ALP79} J.-C. Adam, G. Laval and D. Pesme, Phys. Rev. Lett. \textbf{%
43}, 1671 (1979).

\bibitem{DoxasCary97} I. Doxas and J. R. Cary, Phys. Plasmas \textbf{4},
2508 (1997).

\bibitem{Laval99} G. Laval and D. Pesme, Plasma Phys. Control. Fusion
\textbf{41}, A239 (1999).

\bibitem{BGK} I. B. Bernstein, J. M. Greene, and M. D. Kruskal, Phys. Rev.
\textbf{108}, 546 (1957).
\end{thebibliography}
\end{document}